\documentclass[longauth]{aa}  
\usepackage[utf8]{inputenc}
\usepackage{graphicx}
\usepackage{xcolor}
\usepackage{changes}
\usepackage{amssymb}

\usepackage{txfonts}
\definecolor{Mygreen}{rgb}{0.00, 0.5, 0.5}
\definecolor{Mypink}{rgb}{1.0, 0.0, 0.5}
\definecolor{Myblue}{rgb}{0.00, 0.2, 0.8}
\definecolor{Myred}{rgb}{0.80, 0.2, 0.0}
\usepackage[breaklinks, citecolor=Myblue, linkcolor=Mygreen, urlcolor=Mygreen, colorlinks=true, debug, baseurl=' ']{hyperref}
\usepackage{orcidlink}
\usepackage[switch]{lineno}
\begin{document} 

\newcommand*{\red}{\textcolor{red}}
\def\addref{\textcolor{blue}{[ref]}}

\def\aj{AJ}%
\def\araa{ARA\&A}%
\def\apj{ApJ}%
\def\apjl{ApJL}%
\def\apjs{ApJS}%
\def\aap{Astron. Astrophys.}%
 \def\aapr{A\&A~Rev.}%
\def\aaps{A\&AS}%
\def\mnras{MNRAS}
\def\ssr{SSRv}
\def\nat{Nature}
\def\jcap{JCAP}

   \title{ZTF SN Ia DR2: Impact of the galaxy cluster environment on the stretch distribution of Type~Ia supernovae}

   \author{Ruppin, F. \inst{\ref{ip2i}}\fnmsep\thanks{florian.ruppin@univ-lyon1.fr} \orcidlink{0000-0002-0955-8954}
        \and Rigault, M. \inst{\ref{ip2i}} \orcidlink{0000-0002-8121-2560}
        \and Ginolin, M.\inst{\ref{ip2i}} \orcidlink{0009-0004-5311-9301} 
        \and Dimitriadis, G. \inst{\ref{dublin}}\orcidlink{0000-0001-9494-179X}
        \and Goobar, A.\inst{\ref{okc}},\orcidlink{0000-0002-4163-4996}
        \and Johansson, J.\inst{\ref{okc}}\orcidlink{0000-0001-5975-290X}
        \and Maguire, K.\inst{\ref{dublin}}\orcidlink{0000-0002-9770-3508}
        \and Nordin, J.\inst{\ref{humbolt}}\orcidlink{0000-0001-8342-6274}
        \and Smith, M. \inst{\ref{ip2i},\ref{lancaster}} \orcidlink{0000-0002-3321-1432} 
        \and Aubert, M.\inst{\ref{clermont}}
        \and Biedermann, J. \inst{\ref{ip2i}}
        \and Copin, Y.\inst{\ref{ip2i}} \orcidlink{0000-0002-5317-7518}
        \and Burgaz, U.\inst{\ref{dublin}}\orcidlink{0000-0003-0126-3999}
        \and Carreres, B.\inst{\ref{marseille},\ref{duke}}\orcidlink{0000-0002-7234-844X}
        \and Feinstein, F. \inst{\ref{marseille}}\orcidlink{0000-0001-5548-3466}
        \and Fouchez, D.\inst{\ref{marseille}}\orcidlink{0000-0002-7496-3796}
        \and M$\mathrm{\ddot{u}}$ller-Bravo, T. E., \inst{\ref{barcelona1},\ref{barcelona2}}\orcidlink{0000-0003-3939-7167}
        \and Galbany, L.\inst{\ref{barcelona1},\ref{barcelona2}}\orcidlink{0000-0002-1296-6887}
        \and Groom, S.~L.\inst{\ref{ipac}}\orcidlink{0000-0001-5668-3507}
        \and Kenworthy, W. D.\inst{\ref{okc}}\orcidlink{0000-0002-5153-5983}
        \and Kim, Y.-L.\inst{\ref{lancaster}}\orcidlink{0000-0002-1031-0796}
        \and Laher, R.~R.\inst{\ref{ipac}}\orcidlink{0000-0003-2451-5482}
        \and Nugent, P.\inst{\ref{lbnl},\ref{ucb}}\orcidlink{0000-0002-3389-0586}
        \and Popovic, B.\inst{\ref{ip2i}}\orcidlink{0000-0002-8012-6978}
        \and Purdum, J.\inst{\ref{caltech}}\orcidlink{0000-0003-1227-3738}
        \and Racine, B. \inst{\ref{marseille}} \orcidlink{0000-0001-8861-3052}
        \and Rosnet, P. \inst{\ref{clermont}}\orcidlink{0000-0002-6099-7565} 
        \and Rosselli, D.\inst{\ref{marseille}}\orcidlink{0000-0001-6839-1421}
        \and Smith, R.\inst{\ref{caltech}}\orcidlink{0000-0001-7062-9726} 
        \and Sollerman, J. \inst{\ref{okc2}} \orcidlink{0000-0003-1546-6615}
        \and Terwel, J. H.\inst{\ref{dublin},\ref{not}}\orcidlink{0000-0001-9834-3439}
        }

   \institute{
   Universite Claude Bernard Lyon 1, CNRS, IP2I Lyon / IN2P3, IMR 5822, F-69622 Villeurbanne, France \label{ip2i} 
   \and
   School of Physics, Trinity College Dublin, College Green, Dublin 2, Ireland
   \label{dublin}
   \and
   The Oskar Klein Centre, Department of Physics, AlbaNova, SE-106 91 Stockholm , Sweden
   \label{okc}
   \and
   Institut für Physik, Humboldt-Universität zu Berlin, Newtonstr. 15, 12489 Berlin, Germany
   \label{humbolt}
   \and 
   Department of Physics, Lancaster University, Lancs LA1 4YB, UK 
   \label{lancaster}
   \and 
    Université Clermont Auvergne, CNRS/IN2P3, LPCA, F-63000 Clermont-Ferrand, France
    \label{clermont} 
   \and 
   Aix Marseille Université, CNRS/IN2P3, CPPM, Marseille, France
    \label{marseille}
    \and   
   Department of Physics, Duke University Durham, NC 27708, USA
   \label{duke}
   \and  
   Institute of Space Sciences (ICE-CSIC), Campus UAB, Carrer de Can Magrans, s/n, E-08193 Barcelona, Spain.
   \label{barcelona1}
   \and
   Institut d'Estudis Espacials de Catalunya (IEEC), 08860 Castelldefels (Barcelona), Spain
   \label{barcelona2}
    \and 
   IPAC, California Institute of Technology, 1200 E. California Blvd, Pasadena, CA 91125, USA
   \label{ipac}
   \and
   Lawrence Berkeley National Laboratory, 1 Cyclotron Road MS 50B-4206, Berkeley, CA, 94720, USA
   \label{lbnl}
   \and
    Department of Astronomy, University of California, Berkeley, 501 Campbell Hall, Berkeley, CA 94720, USA
    \label{ucb}
    \and
   Caltech Optical Observatories, California Institute of Technology, Pasadena, CA 91125, USA
   \label{caltech}
   \and
   The Oskar Klein Centre, Department of Astronomy, AlbaNova, SE-106 91 Stockholm , Sweden
   \label{okc2}
      \and 
   Nordic Optical Telescope, Rambla José Ana Fernández Pérez 7, ES-38711 Breña Baja, Spain
   \label{not}
             }


 
  \abstract
   {Understanding the impact of the astrophysical environment on Type Ia supernova (SN Ia) properties is crucial to minimize systematic uncertainties in cosmological analyses based on this probe.}
   {We investigate the dependence of the SN Ia SALT2.4 light-curve stretch on the distance from their nearest galaxy cluster to study the potential effect of the intracluster medium (ICM) environment on the intrinsic properties of SN Ia. }
   {We used the largest SN Ia sample to date and cross-matched it with existing X-ray, Sunyaev-Zel'dovich, and optical cluster catalogs in order to study the relation between the stretch and the distance to the nearest detected cluster from each SN Ia. We modeled the underlying stretch distribution with a Gaussian mixture with relative amplitudes that depended on redshift and clustercentric distance.}
   {We find that the fit quality of the stretch distribution improves significantly when we included the distance-dependant term in the model with a variation of the Akaike information criterion $\rm{\Delta AIC} = -10.2$. Because of the known correlation between galaxy age and distance from the cluster center, this supports previous evidence that the age of the stellar population is the underlying driver of the bimodal shape of the SN Ia stretch distribution. We further computed the evolution of the fraction of quenched galaxies as a function of distance with respect to cluster center from our best-fit model of the SNe Ia stretch distribution and compared it to previous results obtained from $H\alpha$ line measurements, optical broadband photometry, and simulations. We find our estimate to be compatible with these results.}
   {The results of this work indicate that SNe Ia searches at high redshift targeted toward clusters to maximize detection probability should be considered with caution as the stretch distribution of the detected sample would be strongly biased toward the old subpopulation of SNe Ia. Furthermore, the effect of the ICM environment on the SN Ia properties appears to be significant from the center of the clusters up to their splashback radius. This is compatible with previous works based on observations and simulations of a galaxy age gradient with respect to clustercentric distance in massive cluster halos. The next generation of large-area surveys will provide an increase of an order of magnitude in the size of SN Ia and cluster catalogs. This will enable us to analyze the impact of cluster mass on the intrinsic properties of SNe Ia and of the fraction of quenched galaxies in the outskirts of clusters in more detail, where direct measurements are challenging.}


   \maketitle
%

\section{Introduction}\label{sec:intro}

\begin{figure*}
\centering
\includegraphics[width=0.55\textwidth]{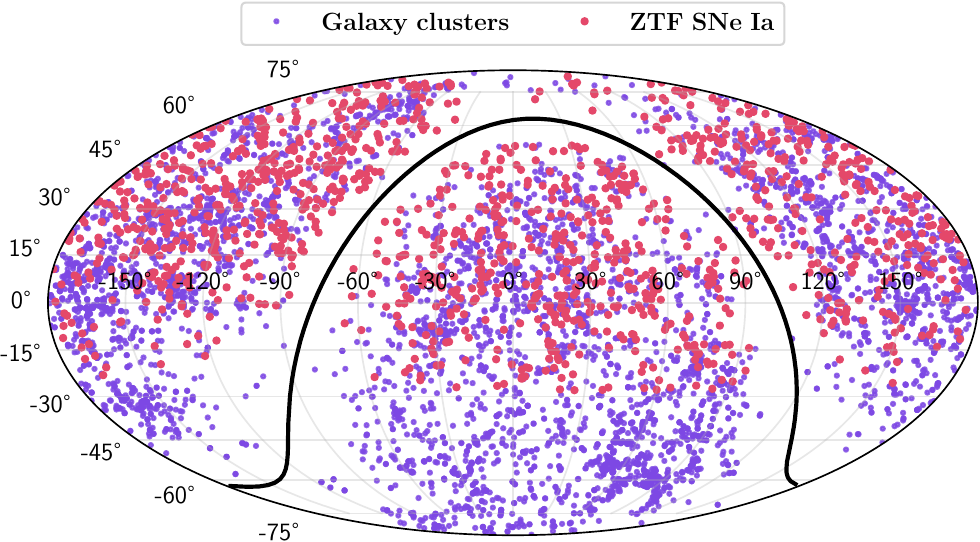}
\hspace{1cm}
\includegraphics[width=0.35\textwidth]{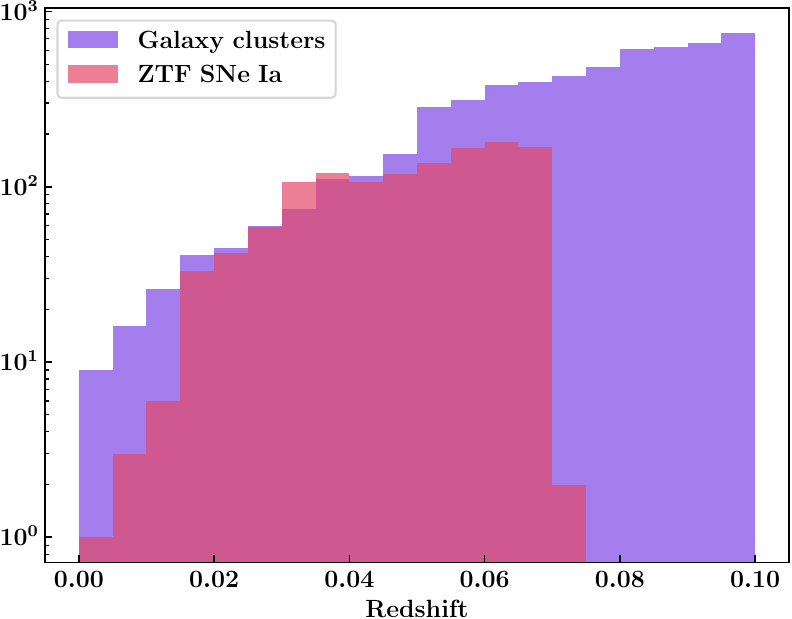}
\caption{Spatial distribution of the SN Ia and cluster samples. \textbf{Left:} Skymap displaying the equatorial coordinates of each object in the cluster (purple) and SN Ia (red) samples. The black line gives the location of the Milky Way in this projection. \textbf{Right:} Redshift distributions of the cluster and SN Ia targets.}
\label{fig:z_space_dist}
\end{figure*}

Type Ia supernovae (SNe Ia) are a key cosmological probe for understanding hitherto unsolved problems such as the root cause of the accelerated expansion of the Universe \citep{rie98,per99}, the discrepancy between late-time and early-time measurements of the Hubble constant \citep[$H_0$; see e.g.][for a review]{div21}, and the validity of general relativity in weak fields at large scales through growth-rate constraints ($f\sigma_8$) at low redshift \citep[e.g.][]{car23}.
In the past two decades, the cosmological quality SN~Ia sample size changed from $\mathcal{O}(100)$ to $\mathcal{O}(2000)$, and the derivation of cosmological parameters from their distances is now {mostly} limited by systematic uncertainties \citep{bet14, sco18,bro22} {although \cite{vin24} realized a cosmological analysis based on SNe Ia in which statistical uncertainties dominated systematics}. Of these, instrumental calibration and astrophysical biases (the limited knowledge of the SN~Ia phenomenon) are the two dominating sources by far. In this paper, we focus on the latter.

With the increase in SN Ia sample size, new effects related to the yet-to-be-understood SN~Ia astrophysical nature have been discovered. Most notably, it has been shown that SNe~Ia from old, red, and/or massive host environments have faster-evolving light curves on average \citep{ham96,how07,nic21} and are $\sim0.1$ mag brighter after the usual stretch and color standardization \citep{tri98, sul10,chi13,kim18,kim19,rig20,smi20,bri22,kel23,rub23}. The SN Ia stretch and color are two properties derived from their light curve that are correlated to their absolute peak brightness in order to reduce the natural scatter of SN Ia brightness of $\sim0.40\,\mathrm{mag}$ to $\sim0.15\,\mathrm{mag}$. {Empirical evidence tends to indicate that the stretch is a property that is correlated to the physics of the progenitor star(s) system}. \cite{nic21} demonstrated that not only is its distribution bimodal, with a low-stretch mode that is only accessible to old-population environments, but data showed that the relative fraction of each mode evolves with redshift, as predicted by the age model from \cite{chi14} and \cite{rig20}. This model is based on earlier observations that two SN~Ia populations must exist to explain the observed SN Ia rates: One population that is associated with recent star formation, called prompt SN Ia, and another population that is associated with stellar mass, called delayed SN Ia \citep[e.g.][]{man05,sul06}. \cite{rig20} modeled the expected evolution of each as a function of redshift based on the evolution of cosmic star formation. \cite{nic21} showed that the low-stretch mode is only accessible to the delayed SN~Ia population. {\cite{wis21} also showed that stretch is related to the delay-time distribution, that is, the rate of SN Ia events as a function of the delay time since the star-formation episode in the host galaxy}.

For SN cosmology, the issue of the environmental dependences in the standardized SN~Ia magnitude is rather troublesome as it may directly lead to biases in the derivation of cosmic distances and thus of cosmological parameters. \cite{rig20} claimed to explain the magnitude offsets in light of the two-population model, prompt and delayed SNe~Ia having their own absolute magnitude, and \cite{bri22} demonstrated that this explanation agrees with existing data. It is not the only possible explanation, however. \cite{bro21} showed that the amplitude of the magnitude offset depends on the SN color. From this observation, they showed that an environment-dependent magnitude-color relation caused by (expected) varying dust properties as a function of the host galaxy properties explains the data well. \cite{pop23} further developed this model, which is now used as reference for the derivation of cosmological parameters from the most advanced SN~Ia compilations \citep{bro22,vin24}. {It is worth noting, however, that \cite{wis22}, \cite{kel23}, and \cite{vin24} showed that a residual step of ${\sim}0.05$~mag is still observed even when the \cite{bro21} dust model is accounted for.} The exact origin and complexity of the SN astrophysical dependences remain to be understood, and this becomes of paramount importance with the vast increase in sample size provided by the current (\textit{Zwicky} Transient Facility, ZTF ; \citealt{bel19a,bel19b,gra19,mas19,dek20}) and future surveys (Large Survey of Space and Time, LSST ; \citealt{lss09}), which will increase the SN~Ia sample size up to $\mathcal{O}(10k)$ and $\mathcal{O}(100k)$, respectively.

In this paper, we thus take a step aside to test a direct prediction from the two-population prompt versus delayed model: Because clusters of galaxies form significantly fewer stars than field galaxies, the distribution of delayed SNe~Ia, and thus, of low-stretch SNe, should be much stronger inside clusters than anywhere else. If this is true, then cosmological surveys targeting clusters at high redshift to boost the SN rate are biased SN searches \citep[see e.g. \emph{Hubble} Space Telescope See Change program][]{hay21}. This bias might be significant if the prompt versus delayed two-population model is indeed the root cause of the aforementioned SN~Ia magnitude biases.


Clusters of galaxies form at the intersections of the cosmic web and represent the largest gravitationally bound structures in the Universe. Their baryonic content is mostly made of a hot diffuse plasma called the intracluster medium (ICM), whose pressure tends to balance the matter infall caused by the gravitational potential that is predominantly produced by the dark matter halo. The late stages of cluster formation are characterized by very energetic events such as mergers with surrounding substructures and mechanical feedback from active galactic nuclei \citep[AGN;][]{rup23}. These events induce an increase in the entropy and temperature of the ICM, produce shock fronts, and promote turbulence in the cluster environment. This thus results in the overall lower star formation rate that is observed in cluster galaxies with respect to their field counterparts, and the typical (low) star formation rate of a galaxy has been shown to depend on its distance from the cluster center \citep[e.g.][]{rak07,rak08,pas19}. 

Studying the SN Ia properties as a function of their location in cluster environments is challenging, however, because only about 10\% of galaxies (and thus of SNe~Ia at first order) are within a cluster \citep{tin08,con16}. Access to a sample of several thousand SNe~Ia is therefore required for any significant signal to be expected. This is only made possible by the new large-area time-domain cosmological surveys such as the ZTF or the Dark Energy Survey \citep[DES;][]{des24}. Nonetheless, a few studies have already been performed on the impact of the cluster environment on SN Ia properties \citep[e.g.][]{toy23}. {Several studies also focused on the rate of SNe Ia in clusters across wide redshift ranges \citep{gra08,mao10,gra14}}. \cite{xav13} identified a ${\sim}3\sigma$ discrepancy between the stretch distributions obtained with 48 SNe Ia detected with the Sloan Digital Sky Survey-II in clusters and 1015 SNe Ia detected in the field. More recently, \cite{lar23} compared the light-curve properties of 102 SNe Ia detected in two different regions of X-ray selected clusters with those of a sample detected in the field. {Finally, \cite{toy23} compared the properties of 66 DES SNe Ia observed in red-sequence-selected galaxy clusters to those of 1024 DES SNe Ia located in field galaxies. None of these three works reported any evidence for variation in the color distributions between the respective subsamples, but they indicated with a high confidence level that the light curves of SNe Ia from clusters evolve faster than those detected in the field, as expected in the \cite{rig20} and \cite{nic21} two population model.}

In this work, we harness the potential offered by the second data release of the ZTF cosmology science working group \citep[DR2;][]{dr2overview}, which contains $\sim3000$ nearby SNe~Ia, for an in-depth study of the SN Ia - galaxy cluster correlations. Instead of dividing our SN Ia sample into different subsamples according to position with respect to known clusters, we propose to study the continuous evolution of the SN Ia light-curve stretch and color parameters with respect to the distance to their nearest detected cluster. Exploiting the fact that low-stretch SNe~Ia only appear to arise from the delayed SN population, we use the SN stretch to probe the fraction of quenched galaxies in clusters as a function of clustercentric distance. We find a remarkably good agreement with previous results based on $H\alpha$ line measurements and optical broadband photometry in cluster galaxies that supports the aforementioned two-population model, and we introduce a novel approach to probing the cluster star formation activity in the context of forthcoming surveys such as the LSST and the High Latitude Time Domain Survey (HLTDS) of the \emph{Roman} Space Telescope \citep[e.g.][]{wan23}.

This paper is organized as follows. In Sect. \ref{sec:samples} we describe our selection method for both the SN Ia and galaxy cluster samples, and we detail the matching procedure that we used in order to determine the nearest cluster to each SN Ia and compute the corresponding distance in Sect. \ref{sec:dist}.  The modeling and the analysis of the environmental drift of the SN Ia stretch distribution as a function of clustercentric distance is presented in Sect. \ref{sec:model} and the results are detailed in Sect. \ref{sec:results}. We interpret and discuss these results in light of the forthcoming SN Ia surveys in Sect. \ref{sec:discuss} before we conclude in Sect. \ref{sec:conclu}. Throughout this work, we adopt a flat $\Lambda$CDM cosmology with $\Omega_m = 0.3$, $\Omega_{\Lambda} = 0.7$, and $H_0 = 70~\mathrm{km\,s^{-1}\,Mpc^{-1}}$.

\begin{figure}
\includegraphics[width=0.5\textwidth]{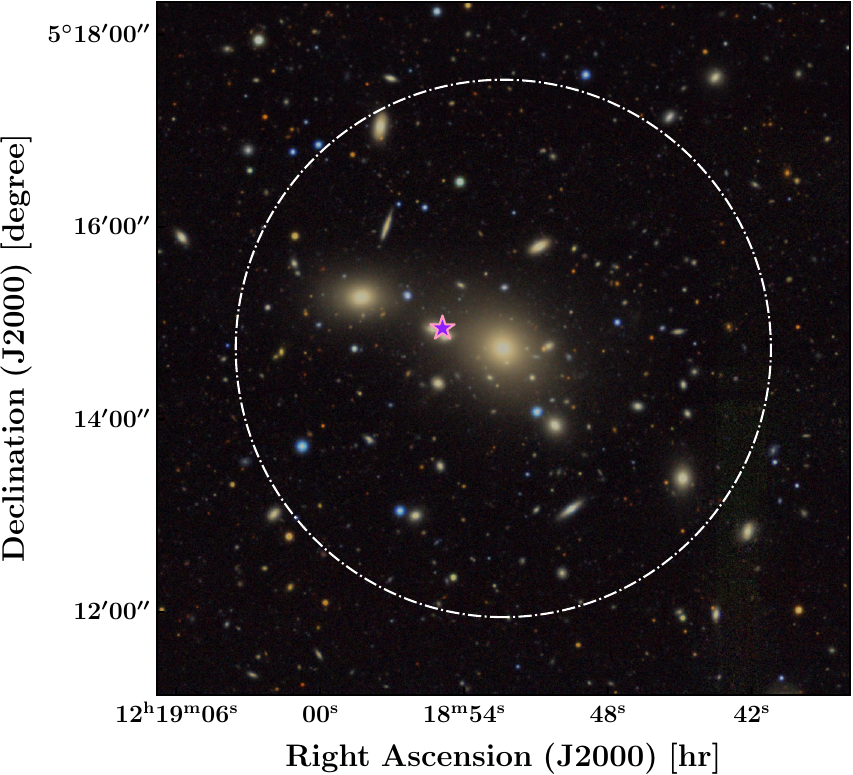}
\caption{{DESI Imaging Legacy Surveys DR10 \citep{dey19} RGB map of the CLJ121852.4+051444 cluster with the ZTF SN Ia ZTF19aacilht (purple star). The white circle is centered on the cluster center and has a radius of $\mathcal{R}_{500}/4$. }}
\label{fig:image}
\end{figure}

\section{Supernova and cluster samples}\label{sec:samples}

This section is dedicated to the presentation of the target selection for the SN Ia and cluster samples. It also describes the matching procedure.

\subsection{ZTF SN Ia sample}

We based our analysis on the second data release of the ZTF cosmology science working group catalog of SNe Ia, $\sim3000$ cosmology grade SNe~Ia \citep[ZTF-Cosmo-DR2 ; ][]{dr2overview}. 
The ZTF survey \citep{bel19a,bel19b,gra19} is acquiring $g$-, $r$-, and $i$-band 47 deg$^2$ images with a typical depth of 20.5 mag in $g$ and $r$ (20 in $i$) and typical two-day cadences (5 in $i$) since March 2018. The ZTF-Cosmo-DR2 sample contains all SNe~Ia that were spectroscopically classified up to December 2020. Most of the spectroscopic classification comes from low-resolution spectroscopy acquired with the spectral energy distribution machine \citep[SEDm,][]{bla18,rig19,kim22}, which is dedicated to classifying ZTF supernovae \citep{fre20,per20}.
The redshifts come either from archival galaxy spectra, mostly from the Sloan Digital Sky Surveys (SDSS), 50\%, or from SN features ($\sim50\%$) and reach a typical $10^{-4}$ and $10^{-3}$ precision, respectively (see \cite{dr2overview} for details).

Following the ZTF-Cosmo-DR2 recommendation \citep{dr2overview}, we used the SN~Ia basic cuts, which ensure sufficient light-curve sampling (seven detections, a color pre- and post-max) and a good estimation of the  SALT2 light-curve parameters ($x_1$, $c$, $t_0$),{ that is, with errors on these parameters given by $\sigma_{x_1} < 1$, $\sigma_c < 0.1$, and $\sigma_{t_0} < 1$ day. Following the work realized by \cite{bur24} on the spectral diversity of ZTF-Cosmo-DR2 SNe Ia, we only kept the SNe Ia that were subclassified} as snia, snia-norm, or snia-pec-91t. {We finally considered a volume-limited cut of $z<0.07$ that is slightly more relaxed than the recommended redshift cut of $z<0.06$. This allowed us to increase our sample size (+351 targets), but did not significantly affect the SALT2 stretch distribution \citep{gin23}, which is the core of our analysis.} We only kept targets with a SALT2 fit-probability $>10^{-4}$ as suggested by \cite{dr2overview}. These cuts left $1249$ targets. We show in Sect. \ref{sec:results} that reducing the redshift cut to $z<0.06$ does not change our conclusions.



\begin{figure*}
\sidecaption
\includegraphics[width=12cm]{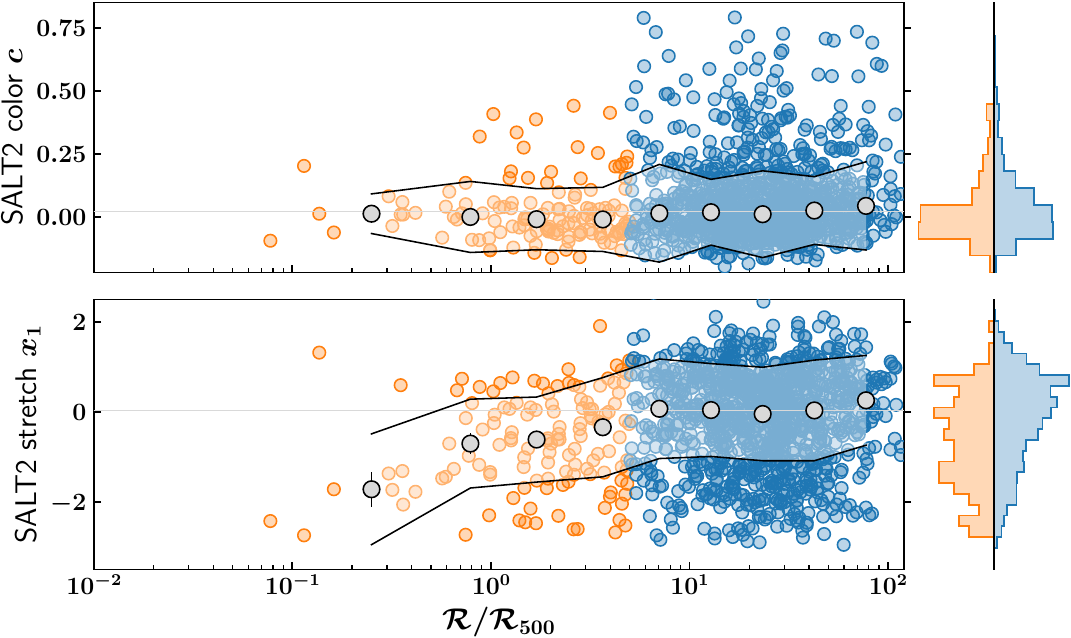}
\caption{SALT2 light-curve color 
(top) and stretch (bottom)
as a function of the scaled projected distance to the nearest galaxy cluster ($\mathcal{R}/\mathcal{R}_\mathrm{500}$). 
We highlight SNe Ia found within (outside) clusters in orange (blue). 
The right part of each panel shows the projected histogram for each subgroup, leftward (rightward) for SNe~Ia within (outside) clusters.
The mean stretch and color values (and scatter) per bin of $\mathcal{R}/\mathcal{R}_\mathrm{500}$ are shown as gray markers (running bands). The error bars associated with these markers are the error on the mean.
}
\label{fig:x1_vs_rr500}
\end{figure*}

\subsection{Galaxy cluster catalogs}

The goal of this work is to study the potential evolution of the SN Ia light curve properties with respect to the distance to the center of their nearest detected cluster. We thus need a cluster sample with spatial and redshift distributions that contain those of our SN Ia sample. There are various ways to detect galaxy clusters. The most robust way is to detect an ICM signature either from its X-ray emission due to thermal bremsstrahlung and line emission \citep[e.g.][]{sep22} or use the inverse-Compton scattering of CMB photons on hot ICM electrons, that is, the Sunyaev-Zel'dovich effect \citep[SZ;][]{sun72,sun80}. 
It is challenging to detect low-mass clusters, however, which are the most abundant clusters, with these techniques because cluster X-ray and SZ emissions directly depend on ICM electron density and temperature, which coevolve with cluster mass. 
Finding clusters through the detection of overdensities of galaxies detected in the optical/infrared (IR) currently leads to much larger samples than are obtained with X-ray and SZ observations. {Depending on the chosen detection algorithm, optical/IR cluster catalogs are in general less pure than their X-ray/SZ counterparts, however, especially at low mass, because of the chance association of field galaxies \citep[e.g.][]{ada19,bul24}.}

We chose to maximize the number of clusters detected in the ZTF survey footprint by considering all available detection channels. In particular, we used the Meta Catalog of X-ray Detected Clusters of Galaxies \citep[MCXC;][]{pif11}, the first cluster catalog of the SRG/eROSITA All-Sky Survey \citep[eRASS;][]{bul24}, the Planck 2nd Sunyaev-Zel'dovich Source Catalog \citep[PSZ2;][]{pla16}, all South Pole Telescope cluster catalogs \citep{ble15,boc19,hua20,ble20}, the latest Atacama Cosmology Telescope cluster catalog \citep{hil21}, and the updated version of the cluster catalog obtained by \cite{wen12} using Sloan Digital Sky Survey (SDSS) III data \citep{wen15}. 
We built our cluster sample from these catalogs by keeping only one occurrence of clusters that were detected by multiple surveys. {Two clusters from different surveys are assumed to be the same object when the two spheres centered on their respective location and with a radius equal to their respective characteristic radius $\mathcal{R}_{500}$ intersect.} This gave us a sample of 5586 galaxy clusters detected at $z < 0.1$. 
The cluster properties that are used in this study are their on-sky position in equatorial coordinates $(\mathrm{R.A.},\mathrm{Dec.})$, their redshift $z$ {which for most of them was obtained from spectroscopic measurements \citep{wen15,bul24}}, and their characteristic radius $\mathcal{R}_{500}$. This radius corresponds to the radius of a sphere with an average density equal to 500 times the critical density of the Universe at the cluster redshift $\rho_c(z)$, such that $M_{500} \equiv 500\rho_c(z)\times\frac{4}{3}\pi \mathcal{R}_{500}^3$. 

As shown in the right panel of Fig.~\ref{fig:z_space_dist}, we extended the redshift range of our cluster sample up to $z = 0.1$ so that we did not miss the nearest clusters of $z{\sim}0.07$ SNe Ia that would happen to be at slightly higher redshifts than the high-end of our SN Ia redshift range.

\subsection{Matching procedure}

The matching procedure we used to find the nearest cluster for each SN Ia in our ZTF sample is similar to the procedure used by \cite{lar23}. We first computed the distance between the SNe Ia and all considered clusters to find the nearest cluster to each SN Ia. To do this, we converted the redshift information into comoving distance using the adopted cosmological model and relied on the 3D matching procedure available in the \texttt{astropy} python library \citep{ast22}. Because clusters are the source of important redshift-space distortions, we chose to also compute the angular separation between each SN Ia and its nearest cluster and used this quantity as the clustercentric distance in our analysis. Following {\cite{xav13}, \cite{toy23}, and \cite{lar23}}, we checked that SNe Ia found at a projected distance lower than $5\mathcal{R}_{500}$ were indeed within the associated cluster by computing a membership probability,
\begin{equation}
p=\frac{1}{\sqrt{2\pi\left(\sigma_{\mathrm{SN}}^{2}+\sigma_{\mathrm{CL}}^{2}\right)}}\int_{-z_{d}}^{+z_{d}}\exp\left[-\frac{(z-[z_{\mathrm{SN}}-z_{\mathrm{CL}}])^{2}}{2\left(\sigma_{\mathrm{SN}}^{2}+\sigma_{\mathrm{CL}}^{2}\right)}\right]\,d z.
\end{equation}
This probability depends on the cluster and SN Ia redshifts, $z_{\mathrm{SN}}$ and $z_{\mathrm{CL}}$, their uncertainties, $\sigma_{\mathrm{SN}}$ and $\sigma_{\mathrm{CL}}$, and on the redshift range $[-z_d,+z_d]$ characterizing the cluster extent along the line of sight in redshift space. We considered that $z_d = 3\sigma_M$, where $\sigma_M$ is the expected velocity dispersion in each cluster given their mass $M_{500}$ estimated using  the \cite{zha11} $\sigma_M - M_{500}$ scaling relation. 
When SNe Ia were found at a projected distance smaller than $5\mathcal{R}_{500}$, we only considered them to be cluster members when $p > 0.9$. When this was not the case, we considered the SN Ia not to be a member of its nearest cluster and its redshift thus to be much less affected by redshift-space distortions. We therefore considered the 3D distance as the clustercentric distance for these rare particular cases; we found only two such cases. Out of the $1249$ SNe Ia in our sample, $131$ are found at a projected distance lower than $5\mathcal{R}_{500}$ from the center of their nearest cluster, and $129$ are cluster members ($p>0.9$). {The two SNe Ia that we lost from our cut on the membership probability can be recovered when we loosen the cut to $p>0.5$, as was done by \cite{xav13} and \cite{toy23}. They are found at $3.9\mathcal{R}_{500}$ and $1.1\mathcal{R}_{500}$ from their alleged host cluster center, respectively. As adding one SN Ia at these respective clustercentric distances does not change the outcome of our work significantly, we preferred to keep our conservative choice on the membership probability to strengthen the purity of the sample at $\mathcal{R}<5\mathcal{R}_{500}$.} 
\begin{figure}
\centering
\includegraphics[width=\linewidth]{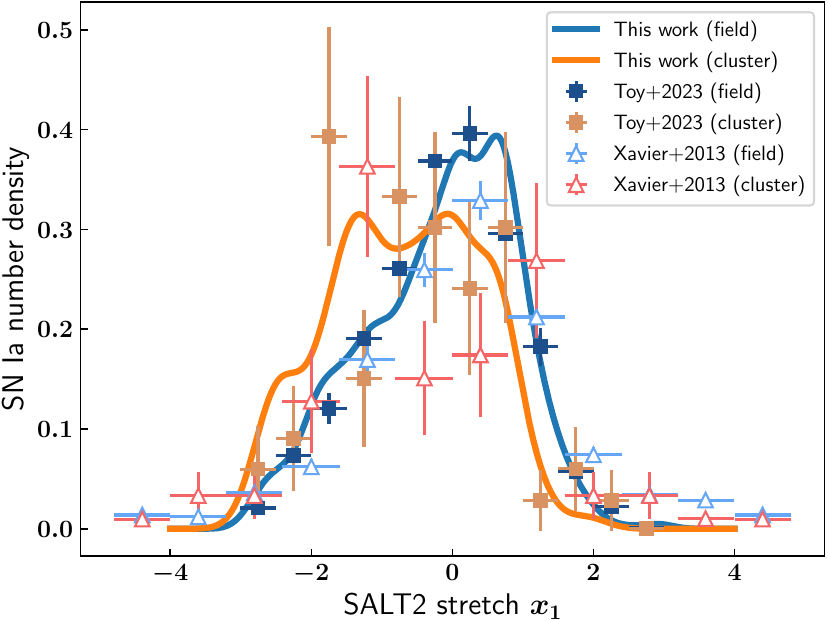}
\caption{Comparison between the stretch distributions observed in clusters ($\mathcal{R} < 5\mathcal{R}_{500}$, orange line) and in the field ($\mathcal{R} > 10\mathcal{R}_{500}$, blue line) with results from previous works. The stretch distributions obtained in clusters (the field) in \cite{xav13} and \cite{toy23} are given by the red-orange (blue) triangles and squares, respectively.
}
\label{fig:comparison}
\end{figure}

We consequently find $\sim10\%$ of our SN~Ia sample to be part of a known cluster. This statistic agrees very well with galaxy surveys, which find that only ${\sim}10\%$ of galaxies at $z < 0.1$ lie within halos with $M_{500} > 10^{13}~\mathrm{M_{\odot}}$, given the halo mass function \citep{tin08}, the galaxy number density in the Universe \citep{con16}, and the cluster mass-richness scaling relation \citep{chi20}. {It is interesting to note that this fraction of SNe Ia found in clusters is consistent with the findings of previous studies that also matched SNe Ia and clusters from well-defined samples, for instance, 6.4\% in \cite{toy23}, 10.1\% in \cite{gra08}, 4.7\% in \cite{xav13}, and 13.8\% in \cite{lar23}.}

We show an example of a ZTF SN Ia found at the core of a galaxy cluster in Fig.~\ref{fig:image}. This DESI Imaging Legacy Surveys DR10 \citep{dey19} RGB map of the CLJ121852.4+051444 cluster is centered on its brightest cluster galaxy (BCG), and we display the ZTF19aacilht SN Ia with a purple star.  It is most probably a member of one of the two galaxies found northeast of the BCG. The dashed white circle indicates the value of $\mathcal{R}_{500}/4$ for this particular cluster, which shows that the ZTF SN Ia is indeed located at the core of this cluster.

\section{SN~Ia properties as a function of distance from their nearest clusters}\label{sec:dist} 

\subsection{SN~Ia properties from cluster-host and field-host galaxies}

The distribution of the SN Ia light-curve stretch and color parameters as a function of the projected clustercentric distance scaled by the projected characteristic radius $\mathcal{R}_{500}$ of their nearest galaxy cluster is shown in Fig.~\ref{fig:x1_vs_rr500}. This figure clearly shows that SNe~Ia that explode within clusters (orange) have lower stretch on average ($x_1$) than SNe Ia from field galaxies ($\Delta \langle x_1 \rangle= -0.51 \pm 0.10$; $5.1\sigma$), while their color seems to be more similar ($\Delta \langle c \rangle= -0.040 \pm 0.011$; $3.6\sigma$). Fig.~\ref{fig:x1_vs_rr500} also shows that the parameter distribution shapes differ, and this is not a simple mean shift. In the stretch, the fraction of low-stretch mode SNe~Ia ($x_1 \lesssim -1$) is proportionally much more populated for cluster SNe~Ia than for those hosted by field galaxies. For the color, SNe~Ia within clusters seem to populate the redder tail less. Both these effects are expected because cluster galaxies host older stars, and thus, a high fraction of delayed SNe~Ia that are characterized by low-stretch light curves. We also expect lower dust contents as galaxies are found closer to the cluster centers. As reported by \cite{mcg10}, dust grains are likely to be destroyed by thermal sputtering in the hot ICM. The increasing density of the ICM toward cluster centers increases the efficiency of this sputtering, which explains why less dust is found in cluster cores than in their outskirts \citep[see][for a review]{shc21}. Then, if dust is indeed the root cause of the redder SN~Ia color tail, we should not observe very reddened SNe~Ia ($c\geq0.2$) within clusters \cite[see e.g.][and the companion paper on SN color \citealt{gin23b}]{jha07,man11,bro21}. 

Qualitatively, the Kolmogorov-Smirnov test provides a p-value of $p=8\times10^{-6}$ that the two stretch distributions arise from the same underlying population, so that this is excluded at the $4.5\sigma$ level. For the color, we find a p-value of $p=4\times10^{-3}$, which means that the two color distribution are unlikely to arise from the same underlying population, but only at the $2.5\sigma$ confidence level. Furthermore, the stretch distribution clearly evolves as a function of their normalized cluster distance, as visible in Fig.~\ref{fig:x1_vs_rr500}, the closer SNe~Ia are to the cluster core, the more they seem to favor the low-stretch mode. This is studied in detail in Sect.~\ref{sec:model}. These results are consistent with the findings of \cite{aub24}, who studied the ZTF DR2 SN Ia properties as a function of the local density in the large-scale structure traced by Voronoi volumes.

{In order to compare the stretch distributions that we obtain within clusters ($\mathcal{R} < 5\mathcal{R}_{500}$) and in the field ($\mathcal{R} > 10\mathcal{R}_{500}$) with those observed in these two environments by previous studies, we display in Fig.~\ref{fig:comparison} the Gaussian kernel density estimation of these stretch distributions found in ZTF (blue and orange lines) with the results from \cite{xav13} and \cite{toy23} (triangles and squares). Given the large uncertainties associated with the stretch distributions of cluster SNe Ia, we find our results to be fully compatible with previous findings in this environment. We note, however, that the amplitude of the low-stretch mode in the stretch distribution of field SNe Ia appears to be slightly lower in the results of \cite{xav13} and \cite{toy23}. This is expected because both works considered a much wider range in SN Ia redshift, and thus, a higher median redshift for the overall SN Ia population. This results in a smaller fraction of SN Ia in old environments \citep{nic21}.}

\begin{figure}
\centering
\includegraphics[width=\linewidth]{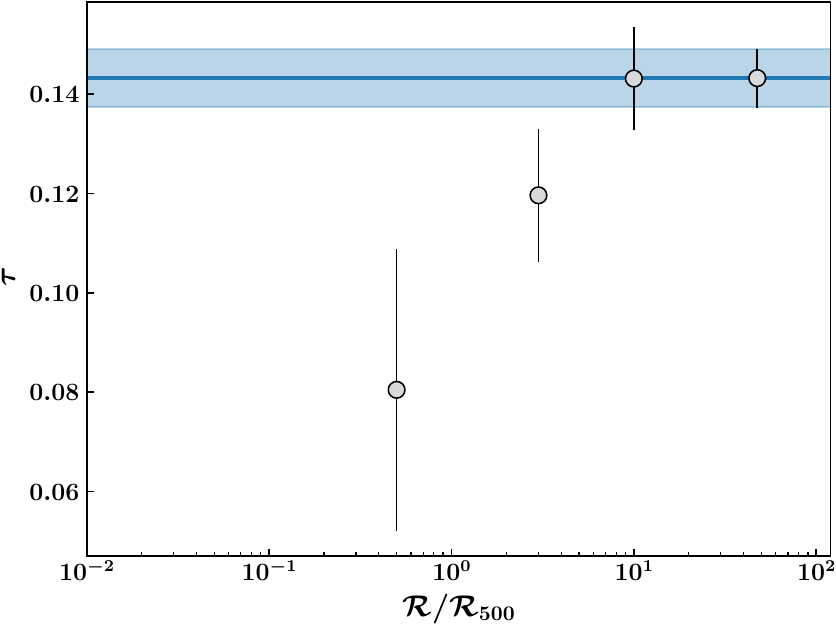}
\caption{Evolution of the dust tail parameter $\tau$ as a function of clustercentric radius. The blue line and area show the best-fit values obtained on the whole sample, and the gray points are obtained in bins of $\mathcal{R}/\mathcal{R}_\mathrm{500}$.
}
\label{fig:tau_vs_rr500}
\end{figure}

{Following \cite{gin23b}, we also modeled the SN Ia color distribution as the convolution of a normal distribution and an exponential decay with a decay parameter $\tau$ in different bins of $\mathcal{R}/\mathcal{R}_\mathrm{500}$. We show the results in Fig.~\ref{fig:tau_vs_rr500}. Although all the fit values (gray points) are compatible with the value obtained on the whole SN Ia sample (blue line) within $2\sigma$, our results show a trend with a decay parameter that decreases as SNe Ia are detected closer to cluster centers.}



\subsection{Effect of the selection function}

It is important to note that our cluster sample is not complete in the considered redshift range. Clusters with $M_{500} \lesssim 10^{14}~\mathrm{M_{\odot}}$ were not all detected at $z < 0.1$. It is therefore likely that some SNe~Ia in our sample might incorrectly not have been associated with a cluster to which they belong. False-positive associations are extremely unlikely, however, because the footprint of clusters in the sky is small.

As a consequence, it is possible that some cluster SNe~Ia are considered as having a field host galaxy. Based on the fraction of galaxies in clusters, however, only a few percent of the field SN sample might arise from a false association like this. The net effect of these targets can thus affect the properties of field-galaxy SN~Ia distributions only very marginally. Conversely, the cluster galaxy sample is likely pure; it might simply have slightly more members if non-false association were made.

Altogether, we do not expect the cluster selection functions to affect our analysis significantly. The SN selection function is only marginal at our fiducial $z<0.07$ redshift cut. Even for higher redshifts, the selection effect can only bias our analysis significantly if the redshifts are significantly different for field- and cluster-host SNe~Ia, which is unlikely.

\section{Modeling the environmental drift of the SN~Ia stretch.}
\label{sec:model} 

This section presents the model we propose to describe the evolution of the SN Ia stretch distribution as a function of scaled clustercentric distance and the analysis procedure we developed to fit our data set. 

\subsection{Stretch distribution}\label{subsec:stretchmod}

Following \cite{rig20} and \cite{nic21}, we assumed that the redshift evolution of the fraction of young-environment SNe Ia (prompt) is given by
\begin{equation}
\delta(z) = \left[K^{-1}\times (1+z)^{-2.8} +  1 \right]^{-1} ,
\end{equation}
where $K = 0.87$ \citep{rig20}. \cite{nic21} show that the SNe~Ia stretch distribution is bimodal, with one mode, the low-stretch one, only available to old-environment (delayed) SNe~Ia. This structure was also confirmed by the companion ZTF-Cosmo-DR2 paper \citep{gin23} and describes the stretch distribution of field galaxies well, as shown in Fig.~\ref{fig:x1_vs_rr500}.


In addition to the known redshift evolution of the stretch distribution, we include the dependence on scaled clustercentric distance $\mathcal{R}/\mathcal{R}_\mathrm{500}$ to the nearest cluster and define the following function to also include the dependence observed in Fig.~\ref{fig:x1_vs_rr500},
\begin{equation}
    \xi(z,\frac{\mathcal{R}}{\mathcal{R}_{500}}) = \left[1 - \left(1+\frac{\mathcal{R} / \mathcal{R}_{500}}{R_{cut}}\right)^{-\gamma}\right] \times \delta(z),
\label{eq:model1}
\end{equation}
with $R_{cut}$ and $\gamma \geq 0$ the new free parameters of this model. This function provides the probability that a given SN~Ia is prompt given its redshift ($\delta(z)$) and its association with a galaxy cluster (the middle term). {The $R_{cut}$ and $\gamma$ parameters give the characteristic radius and the slope of the power-law function that describes the evolution of this probability as a function of clustercentric distance at a particular redshift.} For high values of $\mathcal{R}/\mathcal{R}_\mathrm{500}$, corresponding to field galaxies that largely dominate the SN host distribution, $\xi(z,\frac{\mathcal{R}}{\mathcal{R}_{500}})$ converges toward $\delta(z)$. For an SN that is located very close to a cluster core, its probability to be young tends toward zero, regardess of its redshift.

Following \cite{nic21} and \cite{gin23}, we thus modeled the SN~Ia stretch distribution $\mathcal{P}(x_1)$ knowing their redshift, the host galaxy mass $M_*$, and their location within the cosmic web as follows: 
\begin{equation}
\begin{split}
    \mathcal{P}(x_1 \,|\, z, \frac{\mathcal{R}}{\mathcal{R}_{500}},M_{\rm{env}}) = \,& \xi(z,\frac{\mathcal{R}}{\mathcal{R}_{500}}) \times \mathcal{N}(\mu_1(M_{\rm{env}}),\sigma_1^2)\,+\\
    & (1-\xi(z,\frac{\mathcal{R}}{\mathcal{R}_{500}})) \times \left[a\,\mathcal{N}(\mu_1(M_{\rm{env}}),\sigma_1^2)\right. +\\
    & \left.(1-a)\,\mathcal{N}(\mu_2(M_{\rm{env}}),\sigma_2^2) \right].
\end{split}
\label{eq:model2}
\end{equation}
Here, $M_{\rm{env}} = \rm{log}(M_*/M_{\odot})$, $\mathcal{N}(\mu_1(M_{\rm{env}}),\sigma_1^2)$ is the high-stretch mode, $\mathcal{N}(\mu_2(M_{\rm{env}}),\sigma_2^2)$ is the low-stretch mode that is only accessible to delayed SNe~Ia, and $a$ is the relative fraction of each stretch mode for these supernovae. The two stretch modes have a mean that depends on the host galaxy mass. Following the model introduced in \cite{gin23}, we set the following linear evolution for these means:
\begin{equation}
\mu_{1,2}(M_{\rm{env}}) = \mu_{1,2}^0 + s\times (M_{\rm{env}} - 10).
\end{equation}
Our model of the conditional stretch distribution of SNe Ia given their known scaled distance to their nearest detected cluster and the mass of their host galaxy thus contains eight free parameters $\mathbf{\theta} \equiv (\mu_1^0,\mu_2^0,s,\sigma_1,\sigma_2,a,R_{cut},\gamma) $.


\subsection{Analysis procedure}\label{subsec:analysis}

\begin{figure}
\centering
\includegraphics[width=\linewidth]{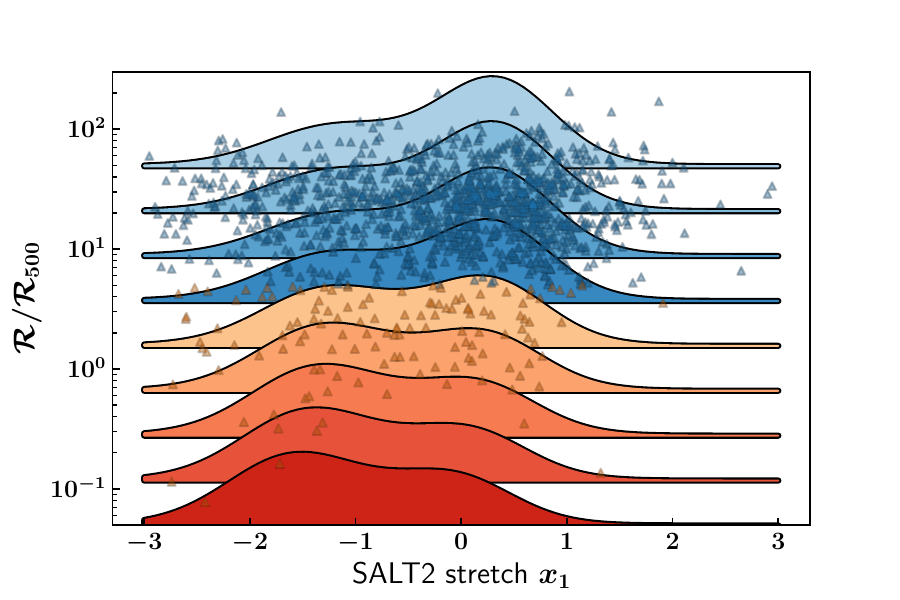}
\caption{{Best-fit conditional distributions of the stretch parameter for different values of $\mathcal{R}/\mathcal{R}_\mathrm{500}$ at the median redshift $\bar{z} = 0.052$. Our data set is shown as orange and blue triangles following the color-code introduced in Fig.~\ref{fig:x1_vs_rr500}.}}
\label{fig:mcmc}
\end{figure}

We assumed that the stretch value of any SN Ia in our sample at a given scaled distance $\frac{\mathcal{R}}{\mathcal{R}_{500}}$ and for a given host mass is a random realization of the conditional distribution $\mathcal{P}(x_1 \,|\, z, \frac{\mathcal{R}}{\mathcal{R}_{500}},M_{\rm{env}})$ defined in Eq. (\ref{eq:model2}). We also included the measurement uncertainty in the distribution so that the overall probability of measuring a given stretch value $x_1^i$ with an error $\Delta x_1^i$ for a given scaled distance $\frac{\mathcal{R}}{\mathcal{R}_{500}}$ is
\begin{equation}
\begin{split}
\mathcal{P}\left(x_1^i \mid \mathbf{\theta} ; \Delta x_1^i, y^i, M_{\rm{env}}^i \right) = \,& y^i \times \mathcal{N} \left(x_1^i \mid \mu_1(M_{\rm{env}}^i),\tilde{\sigma}_{1}^2\right) +\\
 & (1-y^i) \times \left[ a\, \mathcal{N} \left(x_1^i \mid \mu_1(M_{\rm{env}}^i),\tilde{\sigma}_{1}^2\right) + \right. \\
 & \left. (1-a) \, \mathcal{N} \left(x_1^i \mid \mu_2(M_{\rm{env}}^i),\tilde{\sigma}_{2}^2 \right) \right] ,
\end{split}
\end{equation}
with $\tilde{\sigma}_{1,2}^2 = \sigma_{1,2}^2+\Delta x_1^{i\,2}$, $\mathbf{\theta}$ the set of parameters considered in our model (see Sect.~\ref{subsec:stretchmod}), and $y^i = \xi(z^i,\frac{\mathcal{R}^i}{\mathcal{R}_{500}^i})$ the probability of the $i^{\mathrm{th}}$ SN Ia to be prompt. We further assumed that all data points were independent, and we neglected (small) errors on redshift and SN-cluster distances for simplicity. 
We thus maximized the following likelihood function $\mathcal{L}$:
\begin{equation}
\mathrm{ln}(\mathcal{L}) = \sum_i \mathrm{ln} \,\mathcal{P}\left(x_1^i \mid \mathbf{\theta} ; \Delta x_1^i, y^i, M_{\rm{env}}^i \right).
\end{equation}

The fitting procedure is based on a Markov chain Monte Carlo (MCMC) method as made available by the \texttt{emcee} Python library \citep{for13}. We used a Gelman-Rubin convergence criterion as well as a threshold on the number of independent samples after burn-in cutoff based on autocorrelation time in order to stop the MCMC sampling. Using 16 Markov chains, we obtained ${\sim}5000$ independent samples of the posterior after $50\,000$ steps and a burn-in cutoff of $10\,000$ steps.

\section{Results}\label{sec:results}

\subsection{Best-fit model and significance}\label{subsec:res-best-fit}
\begin{table}
\centering
\small
\caption{Priors and best-fit parameters for $\mathcal{P}(x_1 \,|\, z, \frac{\mathcal{R}}{\mathcal{R}_{500}},M_{\rm{env}})$ from Eq~(\ref{eq:model2}).}
\begin{tabular}{r c c} 
\hline\\[-0.8em]
\hline\\[-0.5em]
Parameter & Prior & Best-fit value\\[0.15em]
\hline\\[-0.5em]
$\mu_1^0$ & $\mu_1^0 > \mu_2^0$ & $0.31\pm0.10$\\[0.30em]
$\mu_2^0$ & $\mu_1^0 > \mu_2^0$ & $-1.21\pm0.19$ \\[0.30em]
$s$ & $]-\infty,+\infty[$ & $-0.50\pm0.06$\\[0.30em]
$\sigma_1$ & $[0,2]$ & $0.53\pm0.06$\\[0.30em]
$\sigma_2$ & $[0,2]$ & $0.59\pm0.11$ \\[0.30em]
$a$ & $[0,1]$ & $0.38\pm0.14$\\[0.30em]
$R_{cut}$ & $[0,200]$ & $88\pm24$\\[0.30em]
$\gamma$ & $[0,50]$ & $26\pm7$\\[0.30em]
\hline
\end{tabular}
\label{tab:param}
\end{table}

The best-fit values and $1\sigma$ confidence intervals associated with each model parameter are shown in Table~\ref{tab:param}. We also show the best-fit conditional stretch distributions obtained for several values of the scaled clustercentric distance at the median redshift $\bar{z} = 0.052$ in Fig.~\ref{fig:mcmc}. The $\mu_1^0$, $\mu_2^0$, $\sigma_1$, $\sigma_2$, $s$, and $a$ parameters are compatible with those obtained by \cite{gin23} in the analysis of the SN Ia standardization parameters using the ZTF DR2 volume-limited sample and with that from \cite{nic21} derived from independent samples. 

Our results show an evolution in the overall shape of the conditional stretch distribution given by Eq.~(\ref{eq:model2}) with a subdominant presence of the high-stretch mode in cluster SNe Ia and a low-stretch mode that fades as SNe Ia are found increasingly farther from their nearest detected cluster. At $\mathcal{R} / \mathcal{R}_{500}\sim0$, we find that ${\sim}70\%$ of our SNe Ia are from the low-stretch mode. This also corresponds to findings from \cite{gin23} for a pure delayed SN~Ia old population (reddest environments).

\indent We also performed the analysis procedure described in \ref{subsec:analysis} with $\gamma = 10^5$ and $R_{cut}=10^{-6}$ so that $\xi(z,\mathcal{R}/\mathcal{R}_{500}) = \delta(z)$ for any value of $\mathcal{R}/\mathcal{R}_{500}$ given numerical precision. This analysis under this null hypothesis was conducted in order to test the significance of the evolution of the stretch distribution as a function of $\mathcal{R}/\mathcal{R}_\mathrm{500}$ in addition to its known evolution as a function of host galaxy mass \citep{gin23}. We computed the variation of the Akaike information criterion \citep[$\rm{\Delta AIC}$,][]{aka74} by comparing the posterior probability distributions for the model parameters, in both cases taking the increased number of parameters into account that describe the distance-dependent term. We find that $\rm{\Delta AIC} = -10.2$. This indicates that the null hypothesis is strongly disfavored \citep{bur04}. We thus detect an evolving relative amplitude of the two modes in the stretch distribution as a function of $\mathcal{R}/\mathcal{R}_\mathrm{500}$ that is not solely due to the evolution of the host mass as a function of $\mathcal{R}/\mathcal{R}_\mathrm{500}$. This indicates that the evolution of the stretch distribution as a function of $\mathcal{R}/\mathcal{R}_\mathrm{500}$ is both correlated to the increased average mass and to the average age of host galaxies as SNe Ia are detected ever closer to cluster centers for a given redshift. 

\begin{figure*}
\sidecaption
\includegraphics[width=12cm]{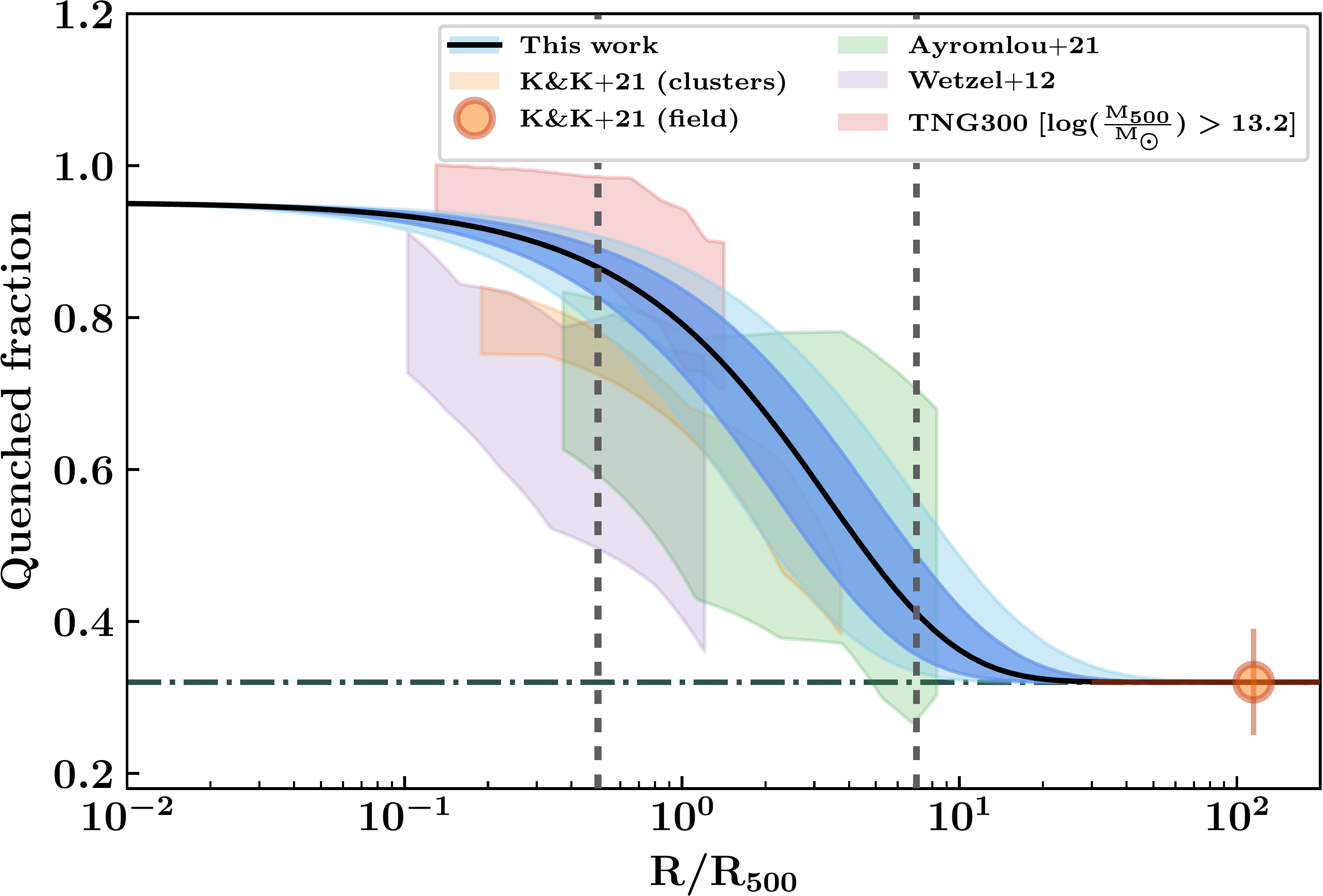}
\caption{{Variations in our estimate of the fraction of quenched galaxies based on the fraction of old SNe Ia given by $1-\xi(z,\frac{\mathcal{R}}{\mathcal{R}_{500}})$ as a function of $\mathcal{R}/\mathcal{R}_\mathrm{500}$ (black line). The $1\sigma$ and $2\sigma$ confidence regions associated with this model are shown in blue. {These error bars are representative of the data constraining power only between the dashed gray lines as the cluster core and field values of the quenched fraction are fixed to literature values.} We compare our results with other estimates of the fraction of quenched galaxies in clusters obtained from $H\alpha$ line measurements \citep[][orange]{kop19}, \citep[][purple]{wet12}, and optical broadband photometry \citep[][green]{ayr21} or with the IllustrisTNG hydrodynamical simulations \citep[][red]{don21}. The fraction of quenched galaxies measured in field galaxies by \cite{kop19} is illustrated as an orange point with error bars. We consider a galaxy to be in the field when $\mathcal{R}/\mathcal{R}_\mathrm{500} > 30$.}}
\label{fig:qfrac}
\end{figure*}

\subsection{Potential selection effects}
We verified that the observed evolution of the SN Ia stretch distribution as a function of $\mathcal{R}/\mathcal{R}_\mathrm{500}$ cannot be caused by selection effects. First, the evolution that we detected indicates that cluster environments favor the explosion of low-stretch SNe Ia, which are known to be fainter than their high-stretch counterparts. A selection effect due to the finite survey sensitivity would thus tend to push for the opposite trend.  Second, we could observe an evolution that is similar to the evolution we detect when the redshift distributions of the SNe Ia found in different bins of $\mathcal{R}/\mathcal{R}_\mathrm{500}$ are different. For example, if most SNe Ia found at $\frac{\mathcal{R}}{\mathcal{R}_{500}} < 5$ lie at very low redshift, but SNe Ia found at $\frac{\mathcal{R}}{\mathcal{R}_{500}} > 10$ are located near the high end of our redshift interval, then we would find a similar trend as the trend displayed in Fig.~\ref{fig:x1_vs_rr500}. However, we find that the SN Ia redshift distributions in the nine bins shown in Fig.~\ref{fig:x1_vs_rr500} show no evidence of a selection-driven trend that might explain our results.\\
The ZTF SN Ia sample is volume-limited up to $z = 0.06$. We verified that our choice of relying on a sample that was not volume-limited in order to maximize sample size cannot induce the observed evolution of the stretch distribution either. We ran the same analysis as presented in Sect.~\ref{sec:model} using a volume-limited sample of SNe Ia by excluding all ZTF SNe Ia found at $z > 0.06$ in our sample. We performed the same analysis also for the 311 SNe Ia that are found at $0.06 < z < 0.07$. The best-fit conditional distributions obtained in both cases are fully compatible, which is consistent with the findings detailed in \cite{gin23}, showing that the stretch distribution of ZTF SNe Ia is representative of the whole population up to $z = 0.07$.

\subsection{Fraction of quenched galaxies}
As discussed in Sect.~\ref{sec:intro}, previous studies have already shown a correlation between the fraction of old galaxies and the clustercentric distance. The results presented in Sect.~\ref{subsec:res-best-fit} are consistent with these previous works because they show that the stretch of SNe Ia found near the center of their nearest cluster has a high probability to be a realization of the low-stretch mode, which has already been shown to be associated with the old galaxy population \citep{nic21,gin23}. Therefore, we propose a new estimate of the fraction of quenched galaxies in clusters $F_{Q}(\mathcal{R}/\mathcal{R}_\mathrm{500})$ based on the fraction of old-environment SNe Ia as a function of $\mathcal{R}/\mathcal{R}_\mathrm{500}$,
\begin{equation}
F_{old}(\mathcal{R}/\mathcal{R}_\mathrm{500}) = 1-\xi(z,\frac{\mathcal{R}}{\mathcal{R}_{500}}).
\label{eq:oldfrac}
\end{equation}
\indent As shown in Eq.~(\ref{eq:model1}), our model leads to $F_{old} = 1$ in the center of clusters. This assumption that all BCGs are old galaxies is valid because numerous studies have shown that the massive elliptical galaxies found at the core of galaxy clusters at low redshift are some of the oldest galaxies in the Universe \cite[e.g.][]{rak07}. The fraction of quenched BCG, that is, BCGs with an $\mathrm{sSFR} < 10^{-11}~\mathrm{yr}^{-1}$, is slightly different, however, because of the feedback mechanisms induced by central AGN that can promote cooling and star formation \citep[e.g.][]{mcd16}. Several follow-up studies of samples of BCG using UV broadband photometry, optical emission-line measurements, and mid-infrared photometry have shown that the fraction of star-forming BCG at $z\sim0$ is about 5\% \citep[e.g.][]{don10,fra14}. {Furthermore, \cite{gra12} found a type II SN Ia in a cluster elliptical galaxy, which is clear evidence that there can be low-level star formation activity in red-sequence galaxies that appear to be quiescent in cluster cores.} We therefore assumed that $F_Q(0) = 0.95$ based on these past studies. We further assumed that the fraction of quenched galaxies obtained for high values of $\mathcal{R}/\mathcal{R}_\mathrm{500}$ is equal to the fraction measured in the field using $H\alpha$ line data, for example, $F_Q^{\mathrm{field}} = 0.32\pm 0.07$ \citep{kop19}. We therefore set $F_Q(100) = 0.32$ because our model is only constrained up to $\mathcal{R}/\mathcal{R}_\mathrm{500} = 100$ (see Fig.~\ref{fig:x1_vs_rr500}). With these assumptions, we estimated the radial variation in $F_Q$ based on Eq.~(\ref{eq:oldfrac}),
\begin{equation}
F_Q(\mathcal{R}/\mathcal{R}_\mathrm{500}) = \alpha \times ( F_{old}(\mathcal{R}/\mathcal{R}_\mathrm{500}) - 1 ) + 0.95,
\end{equation}
with $\alpha = (0.95 - 0.32)/(1 - F_{old}(100))$ the scaling factor used to constrain the value of $F_Q$ at $\mathcal{R}/\mathcal{R}_\mathrm{500} = 0$ and $\mathcal{R}/\mathcal{R}_\mathrm{500} = 100$. We used the median redshift of our volume-limited sample $\bar{z} = 0.052$ to compute this function. We used the converged Markov chains obtained in Sect.~\ref{sec:results} in order to estimate the uncertainty associated with $F_Q(\mathcal{R}/\mathcal{R}_\mathrm{500})$. We represent the best-fit estimate of $F_Q(\mathcal{R}/\mathcal{R}_\mathrm{500})$ as a black line in Fig.~\ref{fig:qfrac}, along with its $1\sigma$ and $2\sigma$ confidence regions in blue. The origin and asymptotic values of this function were set by construction, but the intermediate values of $F_Q$ are fully dependent on the variations of the $\mathcal{P}(x_1 \,|\, z, \frac{\mathcal{R}}{\mathcal{R}_{500}},M_{\rm{env}})$ best-fit model. This model is thus informative only at $0.5\,\mathcal{R}_\mathrm{500} < \mathcal{R} < 7\,\mathcal{R}_\mathrm{500}$ {(between the dashed gray lines)} where ZTF data can constrain $\xi(z,\frac{\mathcal{R}}{\mathcal{R}_{500}})$ and where the confidence regions are not strongly affected by the fixed asymptotic values of $F_Q$.\\
\indent We compared our estimate of the $F_Q(\mathcal{R}/\mathcal{R}_\mathrm{500})$ function with the results obtained by \cite{kop19} by studying the star formation rate of cluster galaxies in $40$ groups and clusters at $0.02 < z < 0.045$ with halo masses comparable to those in our cluster sample (orange region in Fig.~\ref{fig:qfrac}). These results were obtained using the SDSS data release 10 catalog. The authors considered that a cluster galaxy was quenched when its specific star formation rate ($\mathrm{sSFR}$) was such that $\mathrm{log\,sSFR} < -10.75$. We also show the results obtained by \cite{ayr21} (green) and \cite{wet12} (purple), who also used SDSS data in order to estimate the variation in the fraction of quenched galaxies as a function of clustercentric distance in massive halos at $z\sim0$. We find that our estimate of $F_Q(\mathcal{R}/\mathcal{R}_\mathrm{500})$ is compatible with these observational constraints despite their slightly lower values of $F_Q$ at $0.1 < \mathcal{R}/\mathcal{R}_\mathrm{500} < 1$. We further compared our results with the quenched fractions obtained in the IllustrisTNG hydrodynamical simulations \citep{don21} in clusters with a halo mass varying between $M_{200} = 10^{13.5}~\mathrm{M_{\odot}}$ and $M_{200} = 10^{15.2}~\mathrm{M_{\odot}}$ at $z = 0$ (red region) so that it matched the halo-mass distribution of our cluster catalog. Our estimate is fully compatible with these results. This confirms that the evolution of the stretch distribution of SNe Ia depends on the host specific star formation rate and follows the evolution of galaxy age. 

\section{Discussion}\label{sec:discuss}
Previous works have demonstrated the effect of the age gradient due to redshift on the stretch distribution \citep[e.g.][]{nic21}. {This may have an impact on cosmology if the width-luminosity relation also depends on stretch because it would bias the standardized luminosity distances computed for the Hubble diagram.} The analysis presented in this paper shows that the age gradient following clustercentric distance also has an impact on $x_1$. We do not expect this effect to induce any cosmological bias when the considered SN Ia sample is drawn randomly in the cosmic web at all redshifts. However, a cosmological bias could arise if a particular redshift bin in the Hubble diagram is populated by a subsample of SNe Ia that are all detected in the same astrophysical environment. For example, maximizing the SN Ia sample size at high redshift in the Hubble diagram from targeted observations in the direction of known clusters \citep[e.g.][]{hay21} induces a systematic effect on the stretch distribution that may bias SN Ia distances. As shown in \cite{gin23}, the relation between Hubble residuals corrected for color is best described with a broken linear model with two distinct values of the standardization parameter $\alpha$ before and after a cutoff stretch of $x_1^0 = -0.7$. Applying the usual standardization procedure with a single value of $\alpha$ for an inhomogeneous sample of SNe Ia would thus bias the estimates of the distance modulus of the high-$z$ SNe Ia detected in clusters because the majority of their stretch values would be lower than $-0.7$ (cf. Sect.~\ref{sec:results}). Mixing SN Ia samples derived from both large-area and targeted searches in a single Hubble diagram thus requires further including this selection effect in the analysis to avoid biasing cosmological parameters \citep[e.g][]{kes09,bro22b}.\\
\indent In addition to a new systematic effect that has to be accounted for in SN Ia cosmological analyses, the results presented in Sect.~\ref{sec:results} are promising because they offer a new avenue for characterizing the star formation rate of galaxies in the cluster outskirts. We expect the latter to be affected by the violent merger events that occur between the splashback radius and the accretion shock radius of clusters, that is, $5\,\mathcal{R}_{500} < \mathcal{R} < 10\,\mathcal{R}_{500}$ \citep{aun21}. Measuring the star formation rate of cluster galaxies in these regions is challenging, however. Identifying cluster galaxies that are to be followed up with expensive spectroscopic observations is difficult in the outskirts of clusters because their number density is comparable to their density observed in the field. It is possible to estimate a galaxy star formation rate from integrated photometry, but these methods rely on assumptions or independent measurements of the star formation history \citep{joh13}. In this regard, using an independent estimate of the $\mathrm{sSFR}$ of cluster galaxies based on the light-curve properties of SNe Ia enables us to further our knowledge of the entanglement between star formation and environment in the outskirts of clusters.\\
\indent Although the results presented in Fig.~\ref{fig:qfrac} seem to indicate that our method of evaluating $F_Q(\mathcal{R}/\mathcal{R}_\mathrm{500})$ is not very competitive with respect to current estimates based on $H\alpha$ line measurements, we expect a significant improvement in the years to come. The size of the confidence region associated with our best-fit estimate of the conditional probability $\mathcal{P}(x_1 \,|\, z, \frac{\mathcal{R}}{\mathcal{R}_{500}},M_{\rm{env}})$ is currently dictated by the limited number of cluster SNe Ia in our sample. Forthcoming surveys such as LSST \citep{lss09} and the HLTDS \citep{wan23} will increase the available number of SNe Ia by a factor of ${\sim}10$, which will consequently reduce the uncertainty on our model by a factor of ${\sim}3$. This improvement will allow us to probe second-order effects such as the expected dip in $\mathrm{sSFR}$ in the vicinity of the accretion shock radius or the link between ICM metallicity and the properties of the SN Ia stretch distribution. Several observations have indeed shown that there is a sharp increase in the ICM metallicity at $\mathcal{R} < 0.1\,\mathcal{R}_{500}$ \citep[e.g.][]{man17,mcd19}, but the current number of available SNe Ia in these cluster regions is too limited to identify any difference with the light-curve properties measured at $0.1\,\mathcal{R}_{500} < \mathcal{R} < \mathcal{R}_{500}$. The huge increase in SN Ia sample size in the forthcoming years will also enable us to study the effects of cluster mass on the light-curve properties of SNe Ia detected in clusters.  We expect the fraction of quenched galaxies in low-mass clusters to vary differently than in massive ones as a function of $\mathcal{R}/\mathcal{R}_\mathrm{500}$ \citep{don21}. The fraction of quenched galaxies in massive halos, which correspond to the culmination of the hierarchical structure formation, is indeed almost flat because cluster galaxies spend a similar amount of time in their group environment. On the other hand, galaxies found in the outskirts of groups or low-mass clusters may never have made a passage through the dense core regions of their host halo, where environmental effects can lower their SFR. We thus expect the profile of the fraction of quenched galaxies in groups and low-mass clusters to fall rapidly to the field value as $\mathcal{R}/\mathcal{R}_\mathrm{500}$ increases. A precise description of the evolution of the stretch distribution as a function of astrophysical environment will therefore require adding the host cluster mass as an additional parameter in the model of $\mathcal{P}(x_1)$.

\section{Conclusions}\label{sec:conclu}

We have presented a study of the drift of the stretch distribution of SNe Ia as a function of distance to the center of their nearest detected cluster. We based our analysis on a subsample of ZTF DR2 SNe Ia \citep{dr2overview} with good light-curve coverage in a redshift range in which the stretch distribution is not affected by selection effects.  The cluster sample we considered was constructed from multiwavelength catalogs in order to maximize its size.  We associated each SN Ia in our sample with the nearest cluster found in our combined catalog and computed the corresponding scaled projected distance to the cluster center $\mathcal{R}/\mathcal{R}_\mathrm{500}$. We find that $129$ of $1249$ SNe Ia in our sample lie at a projected distance shorter than $5\,\mathcal{R}_{500}$ from their nearest cluster center.\\
\indent Following previous work on the redshift drift of the SN Ia stretch distribution \citep{nic21}, we introduced a model of the conditional distribution $\mathcal{P}(x_1 \,|\, z, \frac{\mathcal{R}}{\mathcal{R}_{500}},M_{\rm{env}})$ that depended on redshift, host galaxy mass, and scaled clustercentric distance.  We also considered a model that did not depend on $\mathcal{R}/\mathcal{R}_\mathrm{500}$ as our null-test. Our conclusions are listed below.
   \begin{enumerate}
      \item The stretch distribution of SNe Ia is significantly dependent on the scaled projected distance to their nearest detected cluster.  The variation in the Akaike information criterion between the new model that we propose and our null-test is $\rm{\Delta AIC} = -10.2$, and we verified that the observed evolution cannot be explained by selection effects alone.
      \item The amplitudes of the two Gaussian modes in our model mostly vary at $0 < \mathcal{R} < 10\,\mathcal{R}_{500}$ for a given redshift.  Our best-fit model indicates that the stretch of the vast majority of SNe~Ia detected at $\mathcal{R}\sim0.1\,\mathcal{R}_{500}$ is a realization of the low-stretch mode. This shows that the environment of the old galaxies found in cluster cores is not suitable for the formation of $x_1 > 0$ SNe Ia.
      \item The evolution of the ratio of the amplitudes of the low-stretch and high-stretch modes in our model as a function of $\mathcal{R}/\mathcal{R}_\mathrm{500}$ is consistent with the galaxy age gradient observed in clusters \citep[e.g.][]{rak07}. This confirms previous findings that the underlying parameter that defines the SN Ia stretch distribution is the age of their host galaxy.
      \item The fraction of delayed SNe Ia in our model of the conditional distribution $\mathcal{P}(x_1 \,|\, z, \frac{\mathcal{R}}{\mathcal{R}_{500}},M_{\rm{env}})$ can be used as a new estimate of the fraction of quenched galaxies.  The evolution of this new estimate as a function of $\mathcal{R}/\mathcal{R}_\mathrm{500}$ can be compared to previous results obtained from spectroscopic measurements of the SFR in cluster galaxies and simulations when we constrain its values in cluster centers and in the field. We find it to be compatible with the radial evolution observed in previous works. This method could open up a promising new avenue of characterizing the link between SFR and environment in the cluster outskirts.
      \item The significant evolution of the SN Ia stretch distribution with clustercentric distance may have repercussions on cosmological SN Ia constraints obtained from inhomogeneous samples in which large-area sky surveys and targeted observations in the direction of known clusters are mixed to maximize the SN Ia detection rate at high redshift. The usual standardization procedure that relies on a single linear model between color-corrected Hubble residuals and stretch would indeed leave a systematic slope in the Hubble residuals of SNe Ia that are detected in clusters \citep{gin23}.
   \end{enumerate}
The forthcoming large-area and fast-cadence surveys LSST and HLTDS will enable a large increase of the number of SNe Ia that are detected in clusters. This will allow us to refine the simple two-parameter power-law model that we propose to describe the evolution of $\mathcal{P}(x_1 \,|\, z, \frac{\mathcal{R}}{\mathcal{R}_{500}},M_{\rm{env}})$ as a function of $\mathcal{R}/\mathcal{R}_\mathrm{500}$. Additional parameters may be added in order to model the expected dip in SFR around the accretion-shock radius of clusters.  The scope of the analysis presented in this work can also be extended by exploring the effect of the cluster mass on the SN Ia stretch distribution and by quantitatively characterizing the effect of this evolution on the cosmological parameters.

\begin{acknowledgements}
Based on observations obtained with the Samuel Oschin Telescope 48-inch and the 60-inch Telescope at the Palomar Observatory as part of the Zwicky Transient Facility project. ZTF is supported by the National Science Foundation under Grants No. AST-1440341 and AST-2034437 and a collaboration including current partners Caltech, IPAC, the Weizmann Institute of Science, the Oskar Klein Center at Stockholm University, the University of Maryland, Deutsches Elektronen-Synchrotron and Humboldt University, the TANGO Consortium of Taiwan, the University of Wisconsin at Milwaukee, Trinity College Dublin, Lawrence Livermore National Laboratories, IN2P3, University of Warwick, Ruhr University Bochum, Northwestern University and former partners the University of Washington, Los Alamos National Laboratories, and Lawrence Berkeley National Laboratories. Operations are conducted by COO, IPAC, and UW.
SED Machine is based upon work supported by the National Science Foundation under Grant No. 1106171
The ZTF forced-photometry service was funded under the Heising-Simons Foundation grant \#12540303 (PI: Graham).
This project has received funding from the European Research Council (ERC) under the European Union's Horizon 2020 research and innovation program (grant agreement n 759194 - USNAC). 
T.E.M.B. acknowledges financial support from the Spanish Ministerio de Ciencia e Innovaci$\mathrm{\acute{o}}$n (MCIN), the Agencia Estatal de Investigaci$\mathrm{\acute{o}}$n (AEI) 10.13039/501100011033, and the European Union Next Generation EU/PRTR funds under the 2021 Juan de la Cierva program FJC2021-047124-I and the PID2020-115253GA-I00 HOSTFLOWS project, from Centro Superior de Investigaciones Cient$\mathrm{\acute{i}}$ficas (CSIC) under the PIE project 20215AT016, and the program Unidad de Excelencia Mar$\mathrm{\acute{i}}$a de Maeztu CEX2020-001058-M.

\end{acknowledgements}

\end{document}